\documentclass[prd,twocolumn,floats,floatfix,nofootinbib,showpacs,superscriptaddress]{revtex4}
\usepackage{amssymb}
\usepackage{citesort}
\usepackage{amsmath,dsfont}
\usepackage{graphicx}
\def\pa{\partial}

\def\al{\alpha}

\def\ii{\textrm i}
\def\ee{\textrm e}

\newcommand{\dbb}{de$\,$Broglie-Bohm}

\newcommand{\bea}{\begin{eqnarray}}
\newcommand{\ena}{\end{eqnarray}}
\newcommand{\be}{\begin{equation}}
\newcommand{\en}{\end{equation}}
\newcommand{\dd}{{\rm d}}

\begin{document}

\title{Wheeler-DeWitt quantization and singularities}

\author{F.~T.~Falciano}\email{ftovar@cbpf.br}
\affiliation{CBPF - Centro Brasileiro de
Pesquisas F\'{\i}sicas, Xavier Sigaud st.\ 150,
zip 22290-180, Rio de Janeiro, Brazil.}

\author{N.~Pinto-Neto}\email{nelsonpn@cbpf.br}
\affiliation{CBPF - Centro Brasileiro de
Pesquisas F\'{\i}sicas, Xavier Sigaud st.\ 150,
zip 22290-180, Rio de Janeiro, Brazil.}

\author{W.\ Struyve}\email{ward.struyve@ulg.ac.be}
\affiliation{Department of Physics, University of Liege, 4000 Liege, Belgium.}

\date{\today}

\begin{abstract}
We consider a Bohmian approach to the Wheeler-DeWitt quantization of the Friedmann-Lema\^itre-Robertson-Walker model and investigate the question whether or not there are singularities, in the sense that the universe reaches zero volume. We find that for generic wave functions (i.e., non-classical wave functions), there is a non-zero probability for a trajectory to be non-singular. This should be contrasted to the consistent histories approach for which it was recently shown by Craig and Singh that there is always a singularity. This result illustrates that the question of singularities depends much on which version of quantum theory one adopts. This was already pointed out by Pinto-Neto {\em et al.}, albeit with a different Bohmian approach. Our current Bohmian approach agrees with the consistent histories approach by Craig and Singh for single-time histories, unlike the one studied earlier by Pinto-Neto {\em et al.} Although the trajectories are usually different in the two Bohmian approach, their qualitative behavior is the same for generic wave functions.
\end{abstract}

\pacs{98.80.Cq, 04.60.Ds, 98.80.Qc}

\maketitle

\section{Introduction}
Recently, some of us investigated the issue of singularities in the Wheeler-DeWitt approach to quantum cosmology and showed that the answer depends strongly on the version of quantum mechanics that one considers \cite{pinto-neto12b}. Two versions were compared: the consistent (or decoherent) histories and the Bohmian approach (also called the \dbb\ or pilot-wave approach). Both approaches have a way of introducing possible histories of the universe. In the Bohmian approach, there is an actual configuration whose time evolution is determined by the wave function. The possible paths of the configuration are the histories. On the other hand, in the consistent histories approach the (coarse-grained) histories are sequences of propositions at different moments in time. (For example, a possible proposition could be that the system occupies a certain region in space).

A simple quantum cosmological model was studied, namely the quantized Friedmann-Lema\^itre-Robertson-Walker model, with a homogeneous scalar field. For the consistent histories approach, Craig and Singh \cite{craig11,craig10} showed that the possible histories are always singular irrespective of the given wave function. Either they start from a singularity or they end in a singularity. On the other hand, according to the Bohmian approach of Pinto-Neto {\em et al.}\ \cite{pinto-neto12b}, there are wave functions for which there are also histories that display a bounce (which corresponds to a universe that contracts until it reaches a minimum volume and then starts expanding) and hence do not have a singularity.

However, while the dynamics of the Bohmian model is very natural, there is no natural probability distribution on the set of histories. As such, there is no immediate way of making probabilistic statements, like about the probability for a history to have a singularity for a particular wave function. On the other hand, Craig and Singh \cite{craig11,craig10} showed that probabilistic statements {\em can} be made in the consistent histories approach. This might seem puzzling since, at least in the context of non-relativistic systems, it is often claimed that consistent histories and Bohmian mechanics yield the same predictions for outcomes of measurements \cite{griffiths99,hartle04}. The reason for this mismatch is that in the consistent histories approach the scalar field is treated (at least formally) as a time variable, whereas it is not in the Bohmian approach. The goal of this paper is to explore another Bohmian dynamics which is also based on treating the scalar field as a time variable. In this case, there is a natural probability distribution for the scale factor which agrees with the one of the consistent histories approach when single-time histories are considered. Hence, this allows for a better comparison between the Bohmian approach and the consistent histories approach concerning the question of singularities. In particular, given a wave function, we can now calculate the probability for a trajectory to run into a singularity. We will find that although the trajectories are different from the ones in the model proposed in \cite{pinto-neto12b}, they are often qualitatively the same. In particular, we will find that some wave functions allow for trajectories that are never singular and instead display a bounce. More precisely, we will see that for a non-classical wave function there is always a non-zero probability for a bounce (while trajectories corresponding to a classical wave function always have a singularity). This should be contrasted with the fact that in the consistent histories approach the probability for a singularity is always one, for any wave function.

The outline of the paper is as follows. We start by recalling the Wheeler-DeWitt quantization of the Friedmann-Lema\^itre-Robertson-Walker model and the Bohmian approach of \cite{pinto-neto12b}. In section \ref{wdwhs}, we will discuss possible choices of Hilbert space that were considered in the consistent histories approach. In section \ref{abm}, we will develop and discuss the alternative Bohmian dynamics. In section \ref{s}, we calculate the probability for a trajectory to be singular. Finally, in section \ref{sch}, we consider the probabilities for singularities in the consistent histories approach, extending the work of Craig and Singh from two-time histories to many-time histories, and compare these to those predicted by the Bohmian approach.

\section{Wheeler-DeWitt quantization and Bohmian mechanics}\label{wdwbm}

A classical flat Friedmann-Lema\^itre-Robertson-Walker space-time is described by a metric
\be
\dd s^2 = N(t)^2 \dd t^2 - \ee^{3\al(t)}  \gamma_{ij}\dd x^i \dd x^j \,,
\label{0.1}
\en
where $N$ is the lapse function, $a=\ee^{\al}$ is the scale factor, and $\gamma_{ij}$ is the flat spatial 3-metric. Considering the matter content described by a free massless homogeneous scalar field $\phi(t)$, the Lagrangian{\footnote{For simplicity, we have dropped the constant factor $4\pi G/3$ in front of $\dot \al^2$. This factor could be removed by a suitable rescaling of the scalar field.}} of the system can be written as \cite{halliwell91b}
\be
L = N \ee^{3\al}  \left( \frac{\dot \phi^2}{2N^2} - \frac{\dot \al^2}{2N^2}\right) \,,
\label{0.2}
\en
where a dot means derivative with respect to coordinate time $t$. The corresponding equations of motion lead to
\be
\dot \phi = \pm \frac{N}{\ee^{ 3\al}} c \,, \qquad \dot \al =   \frac{N}{\ee^{ 3\al}} c \,,
\label{0.3}
\en
where $c$ is an integration constant. $N$ remains an arbitrary function of time. This implies that the dynamics is time reparameterization invariant. Different choices of $N$ merely correspond to different choices of the parameterization of the paths in $(\phi,\al)$ space. In the case $c=0$, the universe is static. For $c \neq 0$, we have
\be
\al = \pm \phi + {\bar c}\,,
\label{0.4}
\en
with ${\bar c}$ another integration constant.

The universe reaches the singularity (i.e., zero volume) when $a=\ee^{\al}=0$. For a non-static universe the singularity is reached for either $\phi \to -\infty$ or $\phi \to \infty$. So the universe either starts with a big bang or ends in a big crunch. In terms of proper time $\tau$, which is defined by $d\tau = N dt$, integration of \eqref{0.3} yields $a=e^\al=\left[3(c\tau + {\tilde c}) \right]^{1/3}$, where ${\tilde c}$ is an integration constant, so that $a=0$ for $\tau = - {\tilde c}/c$ (and there is a big bang if $c>o$ and a big crunch if $c<0$). This means that the universe reaches the singularity in finite proper time.

Canonical quantization of this theory leads to the Wheeler-DeWitt equation
\be
\pa^2_\phi \psi - \pa^2_\al \psi = 0 \,.
\label{1}
\en
In the corresponding Bohmian theory \cite{pinto-neto12b}, there is an actual scalar field $\phi$ and scale factor $a=\ee^{\al}$, which satisfy
\be
\dot \phi = \frac{N}{e^{3\alpha}} \pa_\phi S \,, \quad \dot \al = - \frac{N}{e^{3\alpha}} \pa_\al S \,,
\label{2}
\en
where $\psi = |\psi| \ee^{\ii S}$. The function $N$ is again the lapse function, which is arbitrary, and, as in the classical case, implies that the dynamics is time reparameterization invariant.{\footnote{Note that the time reparameterization invariance is a special feature of Bohmian approaches to mini-superspace models \cite{acacio98,falciano01}. For the usual formulation of Bohmian dynamics for the full Wheeler-DeWitt theory of quantum gravity, a particular space-like foliation of space-time or, equivalently, a particular choice of ``initial'' space-like hypersurface and lapse function, needs to be introduced. Different foliations (or lapse functions) yield different Bohmian theories \cite{shtanov96,pinto-neto99,santini00,pinto-neto02}. So, in this case, the dynamics is not invariant under space-time diffeomorphisms. (Yet, while the usual Bohmian formulation is not diffeomorphism invariant, this may perhaps be achieved with alternative approaches. For a discussion of analogous issues concerning special relativity in Bohmian mechanics, see \cite{duerr14}.)}}

The Wheeler-De Witt equation implies
\be
(\pa_\phi S)^2 - (\pa_\al S)^2 + Q = 0 \,,
\label{3}
\en
\be
\pa_\phi \left( |\psi|^2 \pa_\phi S  \right) - \pa_\alpha \left( |\psi|^2 \pa_\al S  \right) = 0 \,,
\label{4}
\en
where
\be
Q = - \frac{1}{|\psi|} \pa^2_\phi|\psi| + \frac{1}{|\psi|} \pa^2_\al|\psi|
\label{5}
\en
is the quantum potential. If $Q=0$, then \eqref{3} implies that $(\pa_\phi S)^2 = (\pa_\al S)^2$. In addition, $\pa_\al S$ is then conserved along a Bohmian trajectory. Hence, in this case, the Bohmian motion is reduced to the classical motion given by \eqref{0.3}.

Equation \eqref{4} is a continuity equation and implies that the Bohmian dynamics preserves $|\psi|^2$. However, since $|\psi|^2$ is not normalizable it can not be straightforwardly used to make statistical predictions. In this paper, we will consider an alternative Bohmian dynamics, which allows for immediate statistical predictions and which allows for a direct comparison with consistent histories approaches discussed in \cite{halliwell01,craig11,craig10}.

\section{Wheeler-DeWitt equation and Hilbert spaces}\label{wdwhs}
So far we have not introduced a Hilbert space for the Wheeler-DeWitt equation \eqref{1}. We will discuss two possible choices of Hilbert space that have appeared in the literature \cite{ashtekar06,ashtekar08,craig11,craig10,halliwell01}. Both are motivated by considering $\phi$ (at least formally) as a time variable. However, as we shall see, for the purpose of assigning histories to the scale factor, either in the consistent histories or in the Bohmian framework, both choices can be considered equivalent, assuming a suitable choice of ``observable'' for the scale factor in each case.

\begin{figure}
\centering
\includegraphics[width=0.45\textwidth]{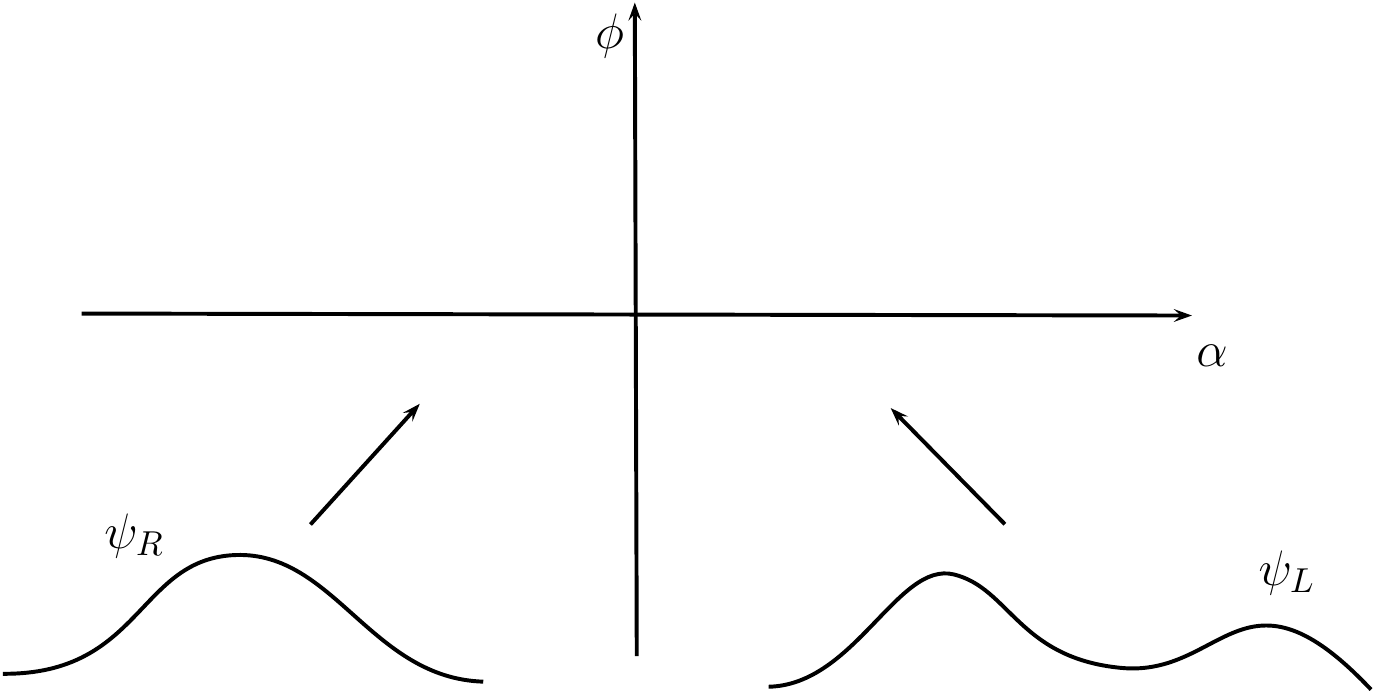}
\caption{\label{fig1}The wave functions $\psi_R$ and $\psi_L$ respectively move to the right and to the left without change in shape.}
\end{figure}

First, consider a general solution to the Wheeler-DeWitt equation \eqref{1}, which is of the form
\be
\psi(\al,\phi)= \psi_L (\al + \phi) + \psi_R (\al - \phi) \,,
\en
where the indices $L$ and $R$ denote respectively ``left-moving" and ``right-moving''. This terminology stems from the fact that if $\phi$ is regarded as time, then $\psi_L$, as a function of $\al$, moves to the ``left'' over time, without change in shape, whereas $\psi_R$ moves to the right, see figure \ref{fig1}.
In terms of the Fourier transform, we have
\begin{align}
\psi(\al,\phi)=& \frac{1}{\sqrt{2\pi}} \int^{\infty}_{-\infty} dk \ \psi_L(k) \ee^{\ii k(\al + \phi)}\nonumber\\
& + \frac{1}{\sqrt{2\pi}} \int^{\infty}_{-\infty} dk \ \psi_R(k) \ee^{\ii k(\al - \phi)} \,.
\end{align}
The general solution can also be decomposed into a positive and negative frequency part
\be
\psi = \psi_+ + \psi_- \,,
\label{}
\en
where
\begin{align}
\psi_+ =& \frac{1}{\sqrt{2\pi}} \int^{\infty}_{0} dk\ \psi_L(k) \ee^{\ii k(\al + \phi)}\nonumber\\
& + \frac{1}{\sqrt{2\pi}} \int^{0}_{-\infty} dk\ \psi_R(k) \ee^{\ii k(\al - \phi)} \,, \\
\psi_- =& \frac{1}{\sqrt{2\pi}} \int^{0}_{-\infty} dk\ \psi_L(k) \ee^{\ii k(\al + \phi)}\nonumber\\
& + \frac{1}{\sqrt{2\pi}} \int^{\infty}_{0} dk\ \psi_R(k) \ee^{\ii k(\al - \phi)} \,,
\end{align}
which satisfy{\footnote{For a wave function $\psi(\alpha) = \frac{1}{2\pi} \int^{\infty}_{-\infty} dk \psi(k) \ee^{\ii k\al}$, the action of the Hamiltonian is defined as $\sqrt{- \pa^2_\al} \psi(\al) = \frac{1}{2\pi} \int^{\infty}_{-\infty} dk |k|\psi(k) \ee^{\ii k\al} $.}}
\be
 \ii \pa_\phi \psi_{\pm}  = {\widehat H}_\pm \psi_{\pm} = \mp \sqrt{- \pa^2_\al} \psi_{\pm}\,.
\label{18}
\en
The left- and right-moving components of $\psi_{\pm}$ will be denoted by $\psi_{L,\pm}$ and $\psi_{R,\pm}$. If one uses the scalar field $\phi$ as a time variable, equation (\ref{18}) represents a Schr\"odinger-like equation and given the initial wave function $\psi^0_{\pm}$ at time $\phi_0$, the (formal) solution is given by
\be
\psi_{\pm}(\alpha,\phi) = \ee^{\pm \ii \sqrt{- \pa^2_\al}(\phi-\phi_0)}\ \psi^0_{\pm}(\alpha) \,.
\label{18b}
\en

The first choice of Hilbert space is ${\mathcal H}_1 = L^2({\mathbb R},{\mathbb C}^2)$ \cite{ashtekar06,craig11,craig10}. We have that ${\mathcal H}_1 = {\mathcal H}_{1,+} \oplus {\mathcal H}_{1,-}$, with ${\mathcal H}_{1,\pm} = L^2({\mathbb R},{\mathbb C})$ the positive and negative frequency Hilbert spaces. We write $\Psi(\al) = (\psi_+(\al),\psi_-(\al))$ with $\Psi \in {\mathcal H}_1$ and $\psi_\pm  \in {\mathcal H}_{1,\pm}$.{\footnote{Actually, in \cite{ashtekar06,craig11,craig10} only the positive frequency sector of ${\mathcal H}_1$ is considered, by appeal to superselection. However, the Bohmian trajectories will depend on whether the wave function is a superposition of both frequencies or not. (Following the terminology of \cite{colin06}, one could say that frequency is weakly superselected, but not strongly.) Therefore, a general analysis should include both frequency sectors.}} The inner product is given by
\be
\langle \Psi | \Phi \rangle_1 = \int d \al \left( \psi^*_+ \phi_+  + \psi^*_- \phi_- \right) \,.
\label{19}
\en
The dynamics for the positive and negative frequency states is given by \eqref{18}. Since the Hamiltonians ${\widehat H}_\pm$ are Hermitian operators on ${\mathcal H}_{1,\pm}$, this dynamics preserves the inner product.

The natural observable corresponding to the scale factor is given by the projection-valued measure (PVM)
\be
{\widehat P}_1(d\al) = {\widehat P}_{1,+}(d\al) \oplus {\widehat P}_{1,-}(d\al)    \,,
\label{19.1}
\en
where
\be
{\widehat P}_{1,\pm}(d\al) = |\al \rangle \langle \al | d\al \,,
\label{19.2}
\en
with $|{\bar \al} \rangle = \delta(\al - {\bar \al})$. (Since $\int {\widehat P}_{1,\pm}(d\al) = {\widehat {\mathds 1}}_{\pm}$, with $ {\widehat {\mathds 1}}_{\pm}$ the identity operator on ${\mathcal H}_{1,\pm}$, it immediately follows that $\int {\widehat P}_{1}(d\al) = {\widehat {\mathds 1}}$, the identity operator on ${\mathcal H}_{1}$.)

For a state $\Psi= (\psi_+,\psi_-)$, the corresponding density for $\al$ is
\be
\rho(\al,\phi) = |\psi_+(\al,\phi)|^2 + |\psi_-(\al,\phi)|^2 \,,
\label{20}
\en
i.e., $\langle \Psi | {\widehat P}_1(d\al)| \Psi \rangle_1 = \rho(\al,\phi)d\al$. (The Hilbert space is the same as that of the Pauli theory for a non-relativistic spin-1/2 particle, with ${\widehat P}_1(d\al)$ being the analogue of the position PVM in the Pauli theory.)

The second choice of Hilbert space ${\mathcal H}_2$ is again the direct sum of positive and negative frequency Hilbert spaces, i.e., ${\mathcal H}_2 = {\mathcal H}_{2,+} \oplus {\mathcal H}_{2,-}$, with ${\mathcal H}_{2,+}={\mathcal H}_{2,-}$ \cite{ashtekar08,halliwell01}. States in ${\mathcal H}_2$ are maps from ${\mathbb R}$ to ${\mathbb C}^2$ and states in ${\mathcal H}_{2,\pm}$  are maps from ${\mathbb R}$ to ${\mathbb C}$. We write ${\bar \Psi}(\al) = ({\bar \psi}_+(\al),{\bar \psi}_-(\al))$ with ${\bar \Psi} \in {\mathcal H}_2$ and ${\bar \psi}_\pm  \in {\mathcal H}_{2,\pm}$ (the bar is used in order to distinguish the states from those in ${\mathcal H}_1$). The inner product on ${\mathcal H}_{2,\pm}$ is given by
\be
\langle {\bar  \psi} | {\bar \phi} \rangle_{2,\pm} = 2 \int d \al \, {\bar \psi}^* (\mp {\widehat H}_\pm) \, {\bar \phi} = 2 \int d \al \, {\bar \psi}^* \sqrt{- \pa^2_\al} \, {\bar \phi} \,,
\label{21}
\en
so that the inner product for ${\mathcal H}_2$ reads{\footnote{Perhaps a more familiar way of writing the inner products \eqref{21} and \eqref{23} is in terms of solutions to the wave equations \eqref{18} \cite{halliwell01}. The inner product \eqref{21} then corresponds to the Klein-Gordon inner product $\langle {\bar \psi} | {\bar \phi} \rangle_{KG} = -\ii \int d\al \left( {\bar \psi}^* \pa_\phi {\bar \phi} - {\bar \phi}  \pa_\phi {\bar \psi}^*\right)$ for two positive frequency solutions to \eqref{18}. The inner product \eqref{23} then corresponds to $\langle {\bar \Psi} | {\bar \Phi} \rangle_2 = \langle {\bar \psi}_+ | {\bar \phi}_+ \rangle_{KG} - \langle {\bar \psi}_- | {\bar \phi}_- \rangle_{KG}$. Note that this inner product is positive definite, unlike the usual definition for the inner product of two solutions to the Klein-Gordon equation, which does not contain the relative minus sign between the positive and negative frequency part.}}
\be
\langle {\bar \Psi} | {\bar \Phi} \rangle_2 = 2 \int d \al \left( {\bar \psi}^*_+ \sqrt{- \pa^2_\al} \, {\bar \phi}_+  + {\bar \psi}^*_- \sqrt{- \pa^2_\al} \, {\bar \phi}_- \right) \,.
\label{23}
\en
The dynamics for the positive and negative frequency states is given again by \eqref{18}. Since the Hamiltonians ${\widehat H}_\pm$ are Hermitian operators on ${\mathcal H}_{2,\pm}$, the dynamics also preserves this inner product.

For the norm of a wave function ${\bar \Psi}$, the integrand on the right hand side of \eqref{23} reads
\be
2\left( {\bar \psi}^*_+ \sqrt{- \pa^2_\al} {\bar \psi}_+  + {\bar \psi}^*_- \sqrt{- \pa^2_\al} {\bar \psi}_- \right) \,.
\label{24}
\en
One might be tempted to regard this quantity as the density of $\al$ at a given time $\phi$ for the state ${\bar \Psi}$. However, it is not always positive definite, even if we restrict ourselves to positive frequencies \cite{kyprianidis85,holland93b}.

One possibility to obtain a positive density is by using the Newton-Wigner states \cite{halliwell01,newton49}
\be
 | {\bar \alpha},NW \rangle  = \frac{1}{2\pi} \int^{\infty}_{-\infty} \frac{dk}{\sqrt{2|k|}}  \ee^{\ii   k( \al -  {\bar \al})} \,.
\label{25}
\en
These states are normalized according to $ \langle \alpha,NW | {\bar \alpha},NW \rangle_2 = \delta(\alpha - {\bar \alpha})$ and form a basis of ${\mathcal H}_{2,\pm}$. The PVM then reads
\be
{\widehat P}_2(d\al) = {\widehat P}_{2,+}(d\al) \oplus{\widehat P}_{2,-}(d\al)   \,,
\label{25.1}
\en
where now
\be
{\widehat P}_{2,\pm}(d\al) =  |\al,NW \rangle  \langle \al ,NW|d\al \,.
\label{25.2}
\en
Note that, as before, $\int {\widehat P}_2(d\al) = {\widehat {\mathds 1}}$, being understood that the action of these operators on the states in the Hilbert space ${\mathcal H}_2$ is through the inner product defined in Eq.~(\ref{23}).

In the consistent histories approach, (coarse-grained) histories for $\al$ can be introduced, by choosing a Hilbert space, a PVM (or positive operator-valued measure (POVM)) and a Hamiltonian. The results will be identical when choosing either $({\mathcal H}_1,{\widehat P}_1,{\widehat H})$ or $({\mathcal H}_2,{\widehat P}_2,{\widehat H})$, where ${\widehat H}$ is the Hamiltonian operator determined by \eqref{18} (the former is considered in \cite{craig11,craig10}, albeit just for positive frequencies, the latter in \cite{halliwell01}).{\footnote{See \cite{goldstein14} for similar considerations on equivalence of quantum theories.}} The reason is that there is a unitary mapping  ${\mathcal H}_2 \to {\mathcal H}_1$, which maps ${\widehat P}_2 \to {\widehat P}_1$ and leaves the Hamiltonian invariant. This mapping is defined by ${\bar \psi}_\pm(\al) \to \psi_\pm (\al) = \langle \al,NW | {\bar \psi}_\pm\rangle_{2,\pm}$ (or $|\al,NW \rangle \to |\al \rangle$ in terms of basis states, or ${\bar \psi}_\pm(k) \to \psi_\pm(k)= \sqrt{2|k|}{\bar \psi}_\pm(k)$ in terms of Fourier components). This map is unitary since $\langle \Psi | \Phi \rangle_1 =\langle {\bar \Psi} | {\bar \Phi} \rangle_2$ (which follows from the fact that $\int {\widehat P}_2(d\al) = {\widehat {\mathds 1}}$, the identity operator on ${\mathcal H}_2$). Since it takes ${\widehat P}_2 \to {\widehat P}_1$, we have
\be
\langle {\bar \Psi} |{\widehat P}_2(d\al)| {\bar \Psi}\rangle_2 = \langle \Psi | {\widehat P}_1(d\al)| \Psi \rangle_1 = \rho(\al,\phi)d\al \,.
\label{28}
\en
As such, we clearly get the same result for probabilities of histories with $({\mathcal H}_1,{\widehat P}_1)$ or $({\mathcal H}_2,{\widehat P}_2)$.

We can also develop a Bohmian approach given a Hilbert space, a POVM and a Hamiltonian. For a given wave function, the distribution corresponding to the PVM will be the Bohmian equilibrium distribution. The dynamics can then be defined in such a way that this distribution is preserved \cite{duerr05a,struyve09a}. In the next section, we will consider a dynamics that preserves the density \eqref{20}. So also in the Bohmian approach, the two choices $({\mathcal H}_1,{\widehat P}_1,{\widehat H})$ and $({\mathcal H}_2,{\widehat P}_2,{\widehat H})$ yield an identical dynamics for the scale factor.

\section{Alternative Bohmian dynamics}\label{abm}
The density $\rho(\al,\phi)$, given in \eqref{20}, is not preserved by the Bohmian dynamics \eqref{2}, i.e., if the density of $\al$ is given by $\rho(\al,\phi_0)$ at a certain time $\phi_0$, then the Bohmian dynamics in general leads to a density different from $\rho(\al,\phi)$ at other times $\phi$ \cite{pinto-neto12b}. However, one can postulate a different dynamics that preserves $\rho$  \cite{berndl96b,goldstein99,struyve09a}. According to this dynamics, possible trajectories $\al(\phi)$ satisfy
\be
\int^{\al(\phi)}_{-\infty} d\bar{\al} \rho(\bar{\al},\phi)  = \int^{\al(0)}_{-\infty} d\bar{\al} \rho(\bar{\al},0) \,,
\label{30}
\en
or, equivalently,
\be
\frac{d\al}{d\phi} = v_2(\al,\phi) = - \frac{1}{\rho(\al,\phi)} \int^{\al}_{-\infty}d\bar{\al} \pa_\phi \rho(\bar{\al},\phi) \,.
\label{31}
\en
Since $\rho$ satisfies the continuity equation
 \be
\pa_\phi \rho + \pa_\al \left(v_2 \rho \right) = 0 \,,
\label{31b}
\en
it is preserved by the Bohmian dynamics.

Let us compare this dynamics to the one formulated in section \ref{wdwbm}. First, consider a real wave function $\psi$, i.e. ${\textrm{Im}} \, \psi=0$.{\footnote{Note that in section \ref{wdwbm}, we wrote the wave function as $\psi = \psi_+ + \psi_-$, whereas in section \ref{wdwhs} we wrote it as $\Psi = (\psi_+,\psi_-)$. These notations are of course equivalent.}} According to the dynamics of section \ref{wdwbm}, the possible solutions are $\al(t)=\al(0),$ $\phi(t)=\phi(0)$, so that the universe is static, with a constant scale factor and constant scalar field. Since these solutions can not be expressed as $\al(\phi)$ they must be different from the trajectories given by \eqref{30}.

Now assume that the trajectories for the first Bohmian dynamics can be expressed as $\al(\phi)$ (at least locally). If $\pa_\phi S \neq 0$, then the vector field tangent to these trajectories is given by
\be
v_1(\al,\phi) = -\frac{\pa_\al S}{\pa_\phi S} \,.
\label{31.2}
\en
The velocity fields $v_1$ and $v_2$ are generically not the same. Using \eqref{3} and \eqref{4}, to evaluate the integral in \eqref{31}, we have that
\begin{align}
v_2 = v_1 - \frac{1}{\rho} \int^{\al}_{-\infty} &d\bar{\al}  \  \frac{\rho}{2(\pa_\phi S)^2} \Big[ \pa_\phi Q   \nonumber\\
& +\frac{2\pa_\phi S}{\rho} \left\{ \pa_\phi \left( \rho \pa_\phi S  \right) - \pa_\al \left( \rho \pa_\al S  \right) \right\} \Big]  \,. 
\label{32}
\end{align}
So a wave function $\psi$ will lead to the same trajectories only if
\be
\pa_\phi Q + \frac{2\pa_\phi S}{\rho} \left\{ \pa_\phi \left( \rho \pa_\phi S  \right) - \pa_\al \left( \rho \pa_\al S  \right) \right\} = 0 \,.
\label{33}
\en

Generically the above condition will not hold. For example, consider a positive or negative frequency wave function. Then the term in curly brackets is zero (because the density $\rho$ then equals $|\psi|^2$ and hence satisfies \eqref{4}) and the condition is reduced to $\pa_\phi Q =0$, which is generically not satisfied.

On the other hand, if the wave function $\psi$ is either only left- or right-moving, i.e., $\psi=\psi_L(\alpha + \phi)$ or $\psi=\psi_R(\alpha - \phi)$, then the condition \eqref{33} is automatically satisfied. In this case, the trajectories are the same for both Bohmian approaches. In addition, the quantum potential \eqref{5} is zero, so that the trajectories are classical. (Conversely, if $Q=0$, then the wave function $\psi$ is either left- or right-moving and the trajectories are the same for both Bohmian approaches.) As shown in section \ref{wdwbm}, the classical trajectories are given by $\alpha = \pm \phi + c$, with $c$ a constant. The positive and negative sign respectively correspond to a right- and left-moving wave function. The trajectories reach the singularity $a=\ee^\al=0$ respectively at $\phi \to -\infty$ and $\phi \to \infty$. Note, however, that in our second Bohmian approach, we have not introduced a lapse function and hence we do not have a straightforward definition of proper time. So, it is not immediately clear whether the singularities are reached in finite proper time or not. Nonetheless, since these trajectories are classical, it seems natural to introduce a proper time that agrees with that of the classical theory. As such, the conclusion is again that the singularities are reached in finite proper time.

This dynamics seems less natural than the one given earlier. The reason is of course that $\phi$ plays the role of time in this dynamics. However, there seems to be no reason to give it that distinguished role. For example, we could equally well have taken $\al$ to play the role of time. This would have resulted in completely different paths. Of course, this is only a simplified model. In a more serious model, we expect that some of the matter fields will represent time on an effective level. But this should follow from analyzing a more fundamental Bohmian dynamics, similar to \eqref{2} (see e.g.\ \cite{shtanov96,goldstein04,pinto-neto05}), rather than be postulated a priori.

\section{Singularities and Bohmian mechanics}\label{s}
In this section, we consider the possibility of singularities (zero scale factor) for trajectories given by the Bohmian dynamics of section \ref{abm} . In particular, using the equilibrium distribution \eqref{20} as probability distribution, we calculate the probability for a trajectory to have a singularity, for a given wave function.

We start by considering the asymptotic behavior of the Bohmian trajectories. First, suppose that the wave function has only right-moving components, i.e., $\Psi(\al,\phi) = \Psi_R(\al - \phi) = (\psi_{R,+}(\al - \phi),\psi_{R,-}(\al - \phi))$. Then the support of the wave function is localized on $\al < 0$ for $\phi \to - \infty$ and $\al > 0$ for $\phi \to + \infty$. This follows from the fact that
\begin{align}
& \lim_{\phi \to - \infty} \int^0_{-\infty}  d\al \rho^{\Psi_R} (\al,\phi) \nonumber\\
&= \lim_{\phi \to - \infty} \int^0_{-\infty} d\al \left( |\psi_{R,+}(\al - \phi)|^2 + |\psi_{R,-}(\al - \phi)|^2 \right) \nonumber\\
&= \lim_{\phi \to - \infty} \int^{-\phi}_{-\infty}  d\nu \left( |\psi_{R,+}(\nu)|^2 + |\psi_{R,-}(\nu)|^2 \right) \nonumber\\
&=|| \Psi_R ||^2_1=1
\end{align}
and, similarly, for the other limit we have
\begin{align}
&\lim_{\phi \to +\infty} \int_0^{\infty}  \dd \al \rho^{\Psi_R} (\al,\phi) \nonumber\\
&= \lim_{\phi \to +\infty} \int_0^{\infty}  \dd \al \left( |\psi_{R,+}(\al - \phi)|^2 + |\psi_{R,-}(\al - \phi)|^2 \right) \nonumber\\
&= \lim_{\phi \to +\infty} \int_{-\phi}^{\infty}  \dd \nu \left( |\psi_{R,+}(\nu)|^2 + |\psi_{R,-}(\nu)|^2 \right) \nonumber\\
&=|| \Psi_R ||^2_1=1 \,.
\end{align}
In the same manner, for a wave function that has only left-moving components, i.e., $\Psi(\al,\phi) = \Psi_L(\al + \phi) = (\psi_{L,+}(\al + \phi),\psi_{L,-}(\al + \phi))$, we have that
\begin{align}
&\lim_{\phi \to -\infty} \int_0^{\infty}  \dd \al \rho^{\Psi_L} (\al,\phi) \nonumber\\
&= \lim_{\phi \to -\infty} \int_0^{\infty}  \dd \al \left( |\psi_{L,+}(\al + \phi)|^2 + |\psi_{L,-}(\al + \phi)|^2 \right) \nonumber\\
&= \lim_{\phi \to -\infty} \int_{\phi}^{\infty}  \dd \nu \left( |\psi_{L,+}(\nu)|^2 + |\psi_{L,-}(\nu)|^2 \right) \nonumber\\
&=|| \Psi_L ||^2_1=1
\end{align}
and
\begin{align}
&\lim_{\phi \to +\infty} \int^0_{-\infty}  d\al \rho^{\Psi_L} (\al,\phi)\nonumber\\
 &= \lim_{\phi \to + \infty} \int^0_{-\infty} d\al \left( |\psi_{L,+}(\al + \phi)|^2 + |\psi_{L,-}(\al + \phi)|^2 \right) \nonumber\\
&= \lim_{\phi \to +\infty} \int^{\phi}_{-\infty}  d\nu \left( |\psi_{L,+}(\nu)|^2 + |\psi_{L,-}(\nu)|^2 \right) \nonumber\\
&=|| \Psi_L ||^2_1=1 \,,
\end{align}
so that the support is now localized on $\al > 0$ for $\phi \to - \infty$ and $\al < 0$ for $\phi \to + \infty$.

So in summary, we have that
\begin{eqnarray*}
\Psi_R\ \text{has support on}\ \begin{cases}
 \al<0\ \text{for}\ \phi \to -\infty \\
\phantom a \\
\al>0\ \text{for}\ \phi \to \infty
 \end{cases} \,,\\
\Psi_L\ \text{has support on}\ \begin{cases}
 \al>0\ \text{for}\ \phi \to -\infty \\
\phantom a \\
\al<0\ \text{for}\ \phi \to \infty
 \end{cases} \,.\\
 \end{eqnarray*}

\begin{figure}
\centering
\includegraphics[width=0.45\textwidth]{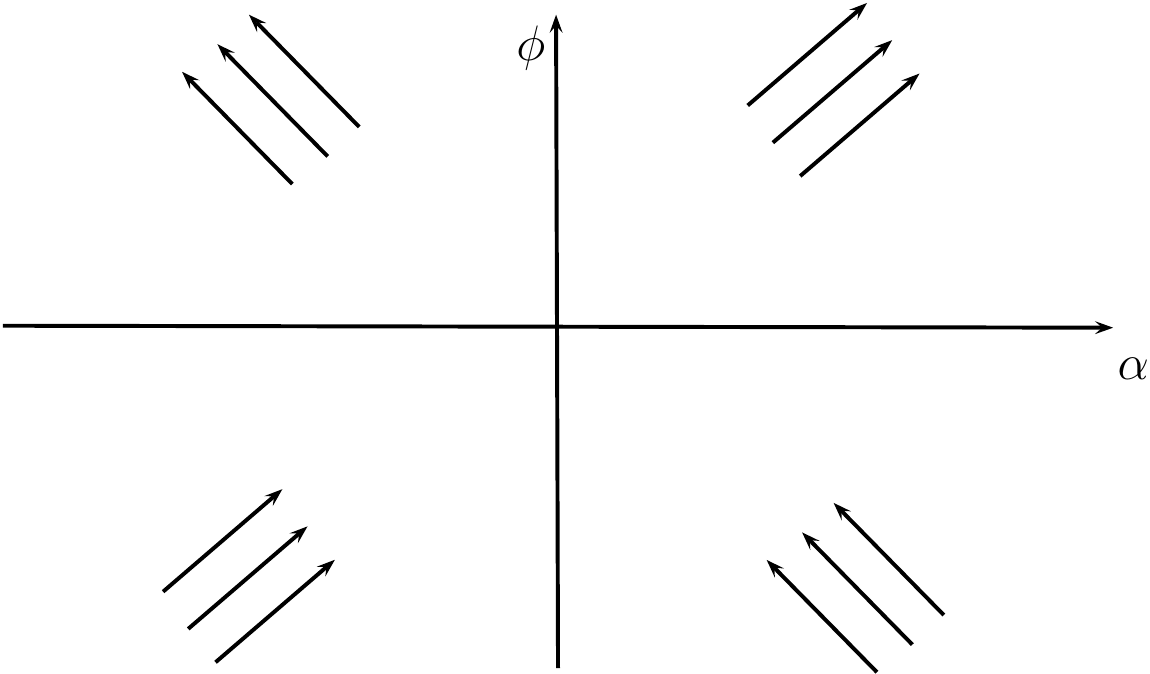}
\caption{\label{fig2} Asymptotic classical behavior of the Bohmian trajectories for an arbitrary wave function.}
\end{figure}

This analysis shows that for any wave function $\Psi = \Psi_R + \Psi_L$, its left- and right-moving components will become asymptotically non-overlapping in $\al$-space. As a consequence, asymptotically, Bohmian trajectories will be determined either by the $\Psi_R$ or $\Psi_L$. Hence, asymptotically, the trajectories are classical, given by $\al = \phi + c$ or $\al = - \phi +c$, see figure \ref{fig2}.\footnote{There might also be a trajectory that does not display this asymptotic behavior and acts as a bifurcation line between trajectories with different possible asymptotic behavior.} (The same holds for the Bohmian dynamics of section \ref{wdwbm}, in the case that the trajectories case be expressed as $\al(\phi)$.)

This asymptotic classical behavior implies that there may be four different types of trajectories. Namely, trajectories that represent a universe that:
\begin{itemize}
\item[1.]
starts with a big bang and keeps expanding forever
\item[2.]
keeps contracting until a big crunch
\item[3.]
starts with a big bang and ends in a big crunch
\item[4.]
undergoes a bounce, i.e., contracts until it reaches a minimal size and then expands again.
\end{itemize}
Only trajectories of type 4 are non-singular. As we shall see, all four types of trajectories may occur. Though, for a given wave function, there can not be trajectories of both types 1 and 2. This is an immediate consequence of the no-crossing property of the Bohmian dynamics (which states that trajectories cannot cross at equal times in configuration space), since such trajectories must cross each other.

In the previous section, we showed that wave functions that are purely left- or right-moving give rise to classical trajectories, which are either of type 1 or type 2, and which are singular. There are also wave functions for which there are trajectories of type 3 or 4.

\begin{figure}
\centering
\includegraphics[width=0.45\textwidth]{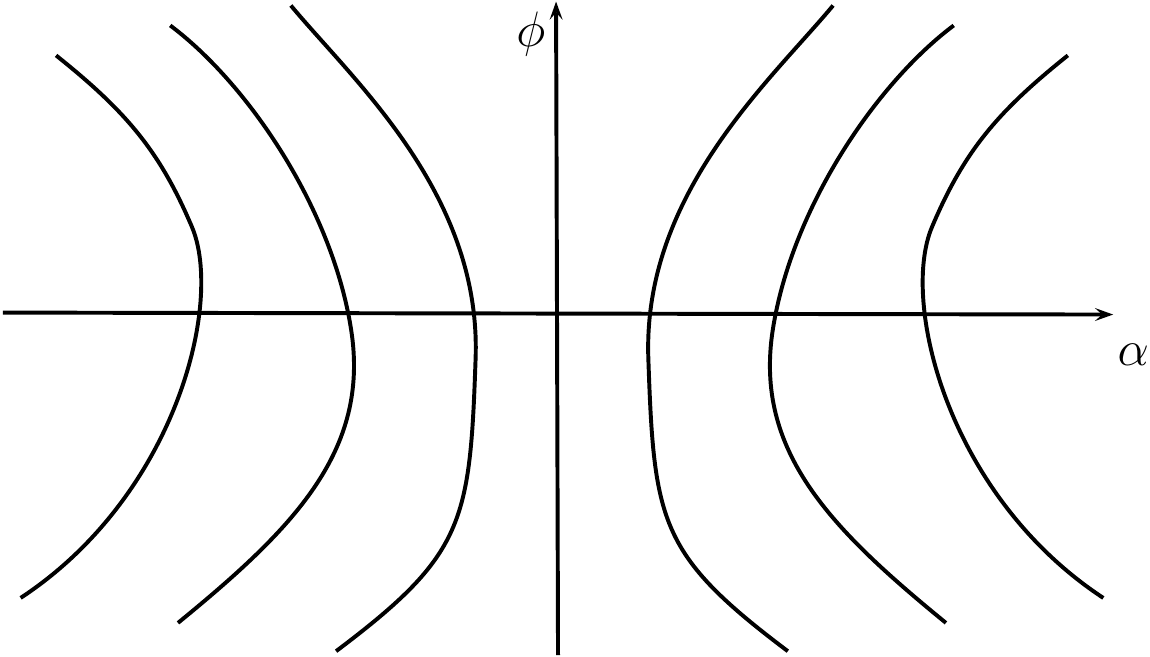}
\caption{\label{fig3} Some trajectories for a wave function that is symmetric under $\alpha \to -\al$. The trajectories on the left represent universes which start with a big bang and end in a big crunch. The trajectories on the right represent universes that bounce.}
\end{figure}

As a simple example, consider a wave function which is even or odd in $\al$. For such a wave function, the velocity field $v_2$ is odd in $\al$. As such, for every possible trajectory $\al(\phi)$, also $-\al(\phi)$ is a possible trajectory. Because of the no-crossing property, this implies that no trajectory will cross the $\phi$-axis.\footnote{Their might also be the trajectory $\al(\phi)=0$ which acts as a bifurcation line between trajectories with different asymptotic behavior.} In other words, trajectories that have positive value of $\al$ at any given time $\phi$, must have $\al >0$ at all times and hence avoid the singularity. These trajectories correspond to a bouncing universe. They asymptotically start with $\alpha = - \phi + c_i$,  reach a minimal value $\al_{min}$ and then asymptotically evolve according to $\alpha =  \phi + c_f$ for $\phi \to \infty$. See figure \ref{fig3} for an example. Conversely, trajectories that asymptotically start with $\alpha = \phi + {\tilde c}_i$ originate from a singularity. Eventually they reach a maximal vale $\al_{max}$ and then for $\phi \to \infty$ they move again to the singularity according to the trajectory $\alpha = - \phi + {\tilde c}_f$. So these trajectories are big bang - big crunch solutions.

Let us now analyze the general case and establish the probability to have singular trajectories for a general wave function $\Psi = \Psi_R + \Psi_L$. Call $P_{L,i}$ and $P_{R,i}$ the probability for the universe to start respectively with $\al < 0$ (from the left) and $\al>0$ (from the right) in the limit $\phi \to -\infty$. Similarly, $P_{L,f}$ and $P_{R,f}$ denote the probabilities for the universe to end up respectively with $\al < 0$ and $\al>0$ in the limit $\phi \to \infty$. Since $\Psi_R$ and $\Psi_L$ do not overlap asymptotically, we have
\begin{align}
P_{L,i} &= \lim_{\phi \to - \infty} \int^0_{-\infty} d\al \rho^{\Psi} (\al,\phi) = ||\Psi_R||^2_1 \,,\nonumber\\
P_{R,i} &= \lim_{\phi \to - \infty} \int^\infty_0 d\al \rho^{\Psi} (\al,\phi) = ||\Psi_L||^2_1   \,,\nonumber\\
P_{L,f} &= \lim_{\phi \to  \infty} \int^0_{-\infty} d\al \rho^{\Psi} (\al,\phi) = ||\Psi_L||^2_1  \,,\nonumber\\
 P_{R,f} &= \lim_{\phi \to  \infty} \int^\infty_0 d\al \rho^{\Psi} (\al,\phi) = ||\Psi_R||^2_1 \,,
\end{align}
and
\be
P_{L,i} = 1 - P_{L,f} = P_{R,f} = 1 - P_{R,i} \,.
\label{50}
\en

We can now find the probability for the universe to start or to end in a singularity for a given wave function $\Psi$. First, assume that $P_{R,i} < P_{R,f}$. In this case, trajectories starting from $\al>0$ as $\phi \to - \infty$ can not end up moving to $\al<0$ as $\phi \to \infty$. If a trajectory {\em did} do that, then all the trajectories that started on the $\al<0$ would must also end up with $\al<0$ as $\phi \to \infty$ because of the no-crossing property. This implies that $P_{L,f} \geqslant P_{L,i}$ and hence, because of \eqref{50}, that $P_{R,i} \geqslant P_{R,f}$, but this contradicts our assumption. Thus, in the case that $P_{R,i} < P_{R,f}$, the probability for a bounce is $P_{\textrm{bounce}} = P_{R,i}$. In addition, it implies that the probability for trajectories to start from a singularity and to keep expanding is $P_{\textrm{expanding}} = P_{R,f} - P_{R,i}$ and the probability for trajectories to start and end in a singularity (i.e., trajectories with a big bang and big crunch) is $P_{\textrm{recollapsing}} = P_{L,i}-(P_{R,f}-P_{R,i})=P_{R,i}$. Finally, the probability for trajectories coming from $\al>0$ to keep contracting towards the singularity is $P_{\textrm{contracting}}= 0$. Similarly, for the case that $P_{R,i} > P_{R,f}$, one can show that $P_{\textrm{bounce}} = P_{R,f}$, $P_{\textrm{contracting}}= P_{L,f}-P_{L,i}$, $P_{\textrm{recollapsing}} =P_{L,i}$ and $P_{\textrm{expanding}} =0$.

In summary, for an arbitrary state $\Psi$, we have
\begin{align}
&P_{\textrm{bounce}} = P_{\textrm{recollapsing}} =  \min (P_{L,i}, P_{L,f}) = \min(P_{R,i}, P_{R,f}) \,,\nonumber\\
&P_{\textrm{expanding}} = \max ( P_{L,i} - P_{L,f} , 0) =  \max ( P_{R,f} - P_{R,i} , 0)\,,\nonumber\\
&P_{\textrm{contracting}}= \max (P_{L,f} - P_{L,i}, 0) =  \max (P_{R,i} - P_{R,f}, 0) \,.
\end{align}
This implies that the probability $P_{\textrm{singularity}} = 1 - P_{\textrm{bounce}}$ to run into a singularity satisfies
\be
\frac{1}{2}  \leqslant P_{\textrm{singularity}} \leqslant 1 \,,
\en
i.e., the probability to run into a singularity is at least one half. Maximum probability is reached when there are only left- or right-moving components. In that case, every trajectory either runs into a singularity or starts from a singularity. On the other hand, for a superposition of left- or right-moving components, there is always a non-zero probability for a bounce. For a state $\Psi$, the maximum probability that can ever be attained for a trajectory to be non-singular is $1/2$, which happens when $P_{L,i}= P_{L,f}$. This happens for the examples of the symmetric and anti-symmetric states discussed before.{\footnote{The example that was used in \cite{griffiths99,hartle04} to compare the Bohmian approach to the consistent histories approach in the context of non-relativistic quantum mechanics is actually completely analogous to the case of a symmetric state in our cosmological model. In \cite{griffiths99,hartle04}, the wave function is composed out of two packets that cross each other in time. The wave function is considered to be completely symmetric so that Bohmian trajectories will never cross the symmetry axis, while the histories in the consistent histories approach correspond to either left- or right-moving paths.\label{g-h}}}

The trajectories for the first Bohmian approach, that was presented in section \ref{wdwbm}, are different from those considered here. However, the trajectories in the first approach that are expressible as $\al(\phi)$ have the same asymptotic behavior as those considered here and hence are qualitatively the same. However, there is no natural probability distribution over trajectories which allows to calculate the probability for trajectories to be singular.

\section{Singularities and consistent histories}\label{sch}
According to the consistent histories approach the probability for a singularity is always one. Coarse-grained histories are either of type 1 or 2. This was shown for two-time histories in \cite{craig11,craig10}, with the times being the infinite past and future.

In \cite{pinto-neto12b}, some of us argued that introducing a third intermediate time would not lead to a family of consistent histories, unless the state is classical (i.e.\ the state is a left- or right-moving wave function). As such, for non-classical states, the consistent histories approach would be unable to deal with the properties of the universe at intermediate times. This would of course undermine the consistent histories approach for this cosmological model.

However, the argument in \cite{pinto-neto12b} is incomplete, mainly because only one particular family of histories was considered. That is, the class of propositions (which are represented by projection operators) was taken to be the same as that in \cite{craig11,craig10} for all three times. And for this class, the histories are not consistent (unless the intermediate time is chosen sufficiently early or late). However, this does not exclude other possible families of histories for which the histories are consistent. We will show that a suitable class of propositions can be chosen such that consistent histories can be considered for an arbitrary number of intermediate times.{\footnote{The analysis is completely analogous to that of \cite{griffiths99,hartle04}. The reason is that, as already noted in footnote \ref{g-h}, we are considering two packets that move across each other, just like in the example of \cite{griffiths99,hartle04}.}} With this choice, the probability for a singularity is always one. This generalizes the results in \cite{craig11,craig10} to an arbitrary number of times.

To setup the consistent histories framework, we will use the triplet $({\mathcal H}_1,{\widehat P}_1,{\widehat H})$. As explained in section \ref{wdwhs}, $({\mathcal H}_2,{\widehat P}_2,{\widehat H})$ would lead to the same results.

A coarse-grained history is a sequence of regions $\Delta_1,\dots,\Delta_n$ in $\alpha$-space at a sequence of times $\phi_1,\dots,\phi_n$. These are obtained by considering an exhaustive set of regions $\{ \Delta^k_{i_k}\}$, $i_k=1,2,\dots$, of $\alpha$-space for each time $\phi_k$, $k=1,\dots,n$.  For each of these regions, there is a projection operator
\be
{\widehat P}^k_{\Delta^k_{i_k}} = \int_{\Delta^k_{i_k}} {\widehat P}_1(d\al)
\en
defined in terms of the PVM ${\widehat P}_1(d\al)$. In terms of these projection operators, the class operator for a history $h=(\Delta^1_{i_1},\dots,\Delta^n_{i_n})$ is
\begin{align}
{\widehat C}_h = {\widehat P}^n_{\Delta^n_{i_n}} {\widehat U}(\phi_n,\phi_{n-1}) {\widehat P}^{n-1}_{\Delta^{n-1}_{i_{n-1}}} {\widehat U}(\phi_{n-1},\phi_{n-2}) \ \cdots  &\nonumber\\
\cdots \  {\widehat P}^2_{\Delta^2_{i_2}} {\widehat U}(\phi_2,\phi_1){\widehat P}^1_{\Delta^1_{i_1}} {\widehat U}(\phi_1,\phi_0) \,, &
\end{align}
where $\phi_0$ is some initial time and ${\widehat U}(\phi_k,\phi_{k-1}) = \ee^{- \ii {\widehat H} (\phi_k - \phi_{k-1} )}$ is the unitary time evolution from time $\phi_{k-1}$ to $\phi_k$.

In order to associate probabilities to a family of histories, the decoherence condition
\be
\langle  {\widehat C}_{h'} \Psi |{\widehat C}_h \Psi \rangle_1 \approx 0 \,, \qquad {\textrm{for }} \quad h' \neq h \,,
\label{100}
\en
needs to be satisfied. The probability for a history $h$ is then given by
\be
P_h = || {\widehat C}_h \Psi||^2_1 \,.
\en

Given a wave function $\Psi$ (and given that $\phi_1$ is sufficiently far in the past or $\phi_n$ sufficiently far in the future), we can always find a collections of sets $\{ \Delta^k_{i_k}\}$ such that the decoherence condition is satisfied. The reason is that the wave function is a superposition of a left-moving and a right-moving packet. For each time $\phi_k$, we can choose the regions $\{ \Delta^k_{i_k}\}$ such that support of $\Psi_L$ and $\Psi_R$ are each approximately within one of the $\Delta^k_{i_k}$ (but not necessarily with the same $i_k$). We then have that for each time $\phi_k$ that ${\widehat P}^k_{\Delta^k_{i_k}} \Psi_L(\alpha,\phi_k) \approx 0$ for all but one $\Delta^k_{i_k}$, which we denote $\Delta^k_L$. For $\Delta^k_L$, we have that ${\widehat P}^k_{\Delta^k_L} \Psi_L(\alpha,\phi_k) \approx \Psi_L(\alpha,\phi_k)$. We then also have that ${\widehat C}_{h_L} \Psi_L (\alpha,\phi_0)\approx \Psi_L (\alpha,\phi_n)$, with $h_L=(\Delta^1_L,\dots, \Delta^n_L)$. In other words, there is a coarse-grained history that approximately follows the track of the packet $\Psi_L$. Similarly, there is a coarse-grained history $h_R$ that approximately follows the track of the packet $\Psi_R$.

If one of the $\phi_k$ is such that $\Psi_L$ and $\Psi_R$ are approximately non-overlapping at that time (for example by taking $\phi_1$ sufficiently early or $\phi_n$ sufficiently late, since we know that the overlap vanishes in the limit $\phi \to \pm \infty$.), then
\begin{align}
{\widehat C}_{h_L} \Psi(\alpha,\phi_0) \approx \Psi_L (\alpha,\phi_n) & \,,&  {\widehat C}_{h_R} \Psi(\alpha,\phi_0) \approx  \Psi_R (\alpha,\phi_n)
\label{100.01}
\end{align}
and
\begin{align}
{\widehat C}_{h} \Psi(\alpha,\phi_0) \approx 0 \ & \,,&   {\textrm{for all}}\quad h\neq h_L,h_R \,.
\end{align}
Since{\footnote{Eq.\ \eqref{101} follows from the fact that a left-moving packet and right-moving packet are orthogonal. Since they have no common support at the far past they are clearly orthogonal then. Since the inner product is preserved over time, they must be orthogonal at all times.}}
\be
\langle {\widehat C}_{h_L} \Psi | {\widehat C}_{h_R} \Psi\rangle \approx 0 \,,
\label{101}
\en
we have that the decoherence condition \eqref{100} is satisfied. The only two coarse-grained histories with non-zero probability are $h_L$ and $h_R$, with probabilities
\begin{align}
P_{h_L} \approx ||  \Psi_L||^2_1 \ &\,,&  P_{h_R} \approx ||\Psi_R||^2_1 \,.
\label{102}
\end{align}

In order to consider the question whether there is a singularity, we should take $\phi_1 \to -\infty$ and $\phi_n \to \infty$. Then \eqref{100.01}-\eqref{102} hold and, using a similar notation as in the previous section, we have
\begin{align}
&P_{\textrm{bounce}} = P_{\textrm{recollapsing}} \approx  0  \,, \nonumber\\
& P_{\textrm{expanding}}  = P_{h_L} \approx  ||\Psi_L||^2_1  \,,\nonumber\\
&P_{\textrm{contracting}}=P_{h_R}  \approx ||\Psi_R||^2_1   \,.
\end{align}

The probability for a bounce is negligibly small. It can actually be made arbitrarily small by suitably choosing the set $\{ \Delta^1_{i_1}\}$. So according to the consistent histories approach, one can always find a consistent family of histories where the probability for a singularity is one.

\section{Conclusion}
We have analyzed a Bohmian approach to the Wheeler-DeWitt quantization of the Friedmann-Lema\^itre-Robertson-Walker model. This Bohmian approach agrees with the consistent histories approach concerning the probabilities for single-time histories. However, it makes different predictions for the probability of trajectories or histories to have a singularity. In the consistent histories approach, at least for the families considered in the present paper, the probability for a history to have a singularity is one. On the other hand, in the Bohmian approach, for generic wave functions (i.e., non-classical wave functions), there is a non-zero probability for a trajectory to be non-singular and have a bounce.

So, as was already emphasized in \cite{pinto-neto12b}, where a different Bohmian model was considered to make the comparison with the consistent histories approach, the question whether or not the Wheeler-DeWitt quantization leads to singularities depends very much on the version of quantum theory one adopts.

In this paper, we have only analyzed the Wheeler-DeWitt quantization. It would be interesting to also study the Bohmian approach to loop quantization. For the consistent histories approach to loop quantization, it was recently shown that histories do not have singularities for generic wave functions \cite{craig13}. It is unclear whether this is also true for a Bohmian approach.

\section*{Acknowledgments}
F.T.F.\ and N.P.-N.\ would like to thank CNPq of Brazil for financial support. W.S.\ acknowledges support from the Actions de Recherches Concert\'ees (ARC) of the Belgium Wallonia-Brussels Federation under contract No.\ 12-17/02. W.S.\ carried out part of this work at Rutgers University, with a grant from the John Templeton Foundation. The opinions expressed in this publication are those of the authors and do not necessarily reflect the views of the John Templeton Foundation. W.S.\ is also grateful to CBPF where part of this work was carried out and to Sheldon Goldstein and Roderich Tumulka for helpful discussions.

\end{document}